\documentclass[prb,twocolumn,superscriptaddress,floatfix,nobalancelastpage,showpacs]{revtex4}

\def\be{\begin{equation}}
\def\ee{\end{equation}}
\usepackage{epsfig}
\usepackage{amsmath}
\usepackage{amssymb}
\usepackage{graphicx}
\usepackage{dcolumn}
\usepackage{multirow}
\newcommand\T{\rule{0pt}{3.1ex}}

\newcommand\B{\rule[-1.7ex]{0pt}{0pt}}

\begin{document} 

\title{Numerical Simulations of the Invar Effect \\ in Fe-Ni, Fe-Pt, and Fe-Pd Ferromagnets}

\author{F. Liot}
\email{f.liot@norinvar.com}
\affiliation{Norinvar, 59 la rue, 50110 Bretteville, France}
\affiliation{Department of Physics, Chemistry and Biology (IFM), Link{\"o}ping University, SE-581 83 Link{\"o}ping, Sweden}

\author{C. A. Hooley}
\affiliation{Scottish Universities Physics Alliance (SUPA), School of Physics and Astronomy, University of St Andrews, North Haugh, St Andrews, Fife KY16 9SS, U.K.}

\date{\today}

\begin{abstract}
The Invar effect in ferromagnetic Fe-Ni, Fe-Pt, and Fe-Pd alloys is investigated theoretically by means of a computationally efficient scheme. The procedure can be divided into two stages: study of magnetism and calculations of structural properties. In the first stage, an Ising model is considered and fractions of Fe moments which point up as a function of temperature are determined. In the second stage, density-functional theory calculations are performed to evaluate free energies of alloys in partially disordered local moment states as a function of lattice constant for various temperatures. Extensive tests of the scheme are carried out by comparing simulation results for thermal expansion coefficients of Fe$_{1-x}$Ni$_{x}$ with $x = 0.35, 0.4,\ldots, 0.8$, Fe$_{0.72}$Pt$_{0.28}$, and Fe$_{0.68}$Pd$_{0.32}$ with measurements. The scheme is found to perform well, at least qualitatively, throughout the whole spectrum of test compounds. For example, the significant reduction of the thermal expansion coefficient of Fe$_{1-x}$Ni$_{x}$ as $x$ decreases from 0.55 to 0.35 near room temperature, which was discovered by Guillaume, is reliably reproduced.
As a result of the overall qualitative agreement between theory and experiment, it appears that the Invar effect in Fe-Ni alloys can be investigated within the same computational framework as Fe-Pt and Fe-Pd.
\end{abstract}


\maketitle

\section{Introduction} \label{sec_intro}

Fe-based materials are used for various technological applications such as springs in watches, car bodies, magnetic cores, and heads of hard disk drives. Despite their ubiquity in everyday life, they exhibit intriguing phenomena that include, among others, high-temperature superconductivity in Fe pnictides \cite{kamihara08}, Fermi-liquid breakdown in Fe-Nb alloys \cite{brando08}, and the Invar effect in transition-metal alloys \cite{wassermann90}. Discovered more than 100 years ago, Invar Fe-based materials display anomalously small thermal expansion coefficients over broad temperature ranges. Fe-Ni alloys with a Ni concentration of about 35 at.\% were the first to be found \cite{guillaume97}. Subsequently, other Invar Fe-based materials were reported, some showing ferromagnetism (e.g., Fe$_{0.68}$Pd$_{0.32}$ \cite{matsui80}) and some antiferromagnetism (e.g., Fe$_{2}$Ti \cite{wassermann98}).

Despite the general consensus that the Invar effect in Fe-based ferromagnets occurs as a result of magnetism, the mechanism giving rise to the Invar phenomenon remains controversial. Two prominent questions raised by recent publications \cite{khmelevskyi05,ruban07} have yet to be answered before the Invar effect is fully understood: (i) Does the anomaly in Fe-Ni appear when changes in the magnitude of local magnetic moments with increasing temperature become anomalously large? (ii) Are the anomalies observed in Fe-Pt, Fe-Pd, and Fe-Ni governed by the same underlying physics?

\begin{figure}
\includegraphics[width=8cm]{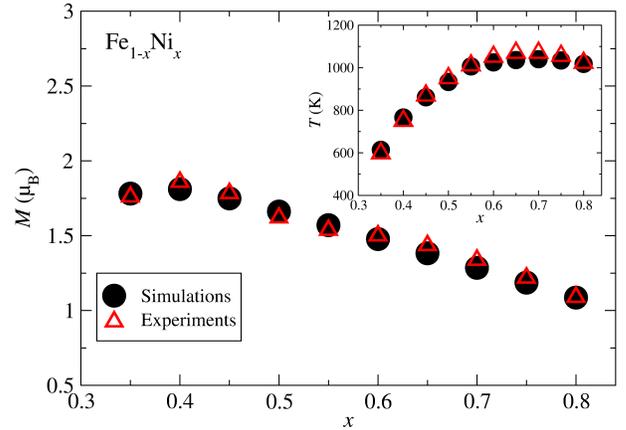}
\caption{Average magnetic moment per atom of fcc Fe$_{1-x}$Ni$_{x}$ at zero temperature plotted against nickel atomic concentration, according to the Ising model (circles) and experiments \cite{crangle62} (triangles). Inset: Concentration dependence of calculated Curie temperature (circles) and measured Curie temperature \cite{crangle62} which has been rescaled by the factor 1.23 (triangles). This figure illustrates step 1 of the numerical method which we have designed to investigate the Invar effect in ferromagnetic Fe-Ni, Fe-Pt, and Fe-Pd.}
\label{figure1}
\end{figure}

Obviously, any unified theory of thermal expansion in Fe-based ferromagnets should capture the Invar effect in ferromagnetic disordered face-centered cubic (fcc) Fe-Ni, Fe-Pt, and Fe-Pd within a single framework. In principle, the linear thermal expansion coefficient of disordered fcc Fe$_{1-x}${\it A}$_{x}$ with {\it A}=Ni, Pt, Pd at zero pressure can be derived from the Helmholtz free energy which depends explicitly on length and temperature. In reality, no applications of density-functional theory (DFT) to {\it ab initio} calculations of finite-temperature free energies have been reported to date. One of the major issues in implementing this strategy is how to incorporate magnetism correctly within current approximations to the exchange and correlation functional \cite{khmelevskyi05, abrikosov07}.

In a recent Letter \cite{khmelevskyi03}, the magnetic contribution to the fractional change in length as a function of temperature was studied theoretically for the case of disordered fcc Fe-Pt. As in our work, the disordered local moment (DLM) formalism \cite{footnote1, gyorffy85, johnson90} was used. However, unlike in our investigation, effects of lattice vibrations on structural quantities were neglected.

The rest of the paper is organized as follows. First, Sec.~\ref{methods I} introduces a scheme to study the temperature dependence of the linear thermal expansion coefficient of ferromagnetic disordered fcc Fe$_{1-x}${\it A}$_{x}$ with {\it A}=Ni, Pt, Pd. A local-moment model is employed to examine magnetic properties; DFT-based calculations and the Debye-Gr{\"u}neisen model \cite{moruzzi88, herper99} provide complementary approaches for determining contributions to free energies. In Sec.~\ref{results I}, the scheme is tested on alloys with different chemical compositions by comparing numerically calculated thermal expansion coefficients with experimental measurements. Finally, Sec.~\ref{conclusion} summarizes our findings. Our work points out the possibility to investigate the Invar effect in Fe-Ni, Fe-Pt, and Fe-Pd ferromagnets within the same computational framework. \\

\begin{figure}
\includegraphics[width=8cm]{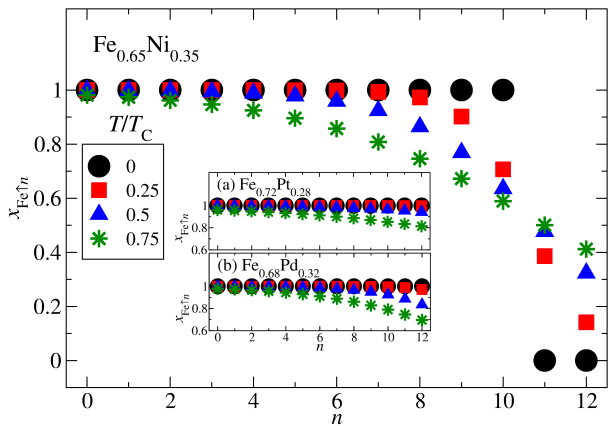}
\caption{Estimated fraction of all Fe moments with $n$ Fe first neighbors which point up in fcc Fe$_{0.65}$Ni$_{0.35}$ plotted against $n$ for temperatures below the Curie temperature, $T_{\rm C}$. Corresponding results for Fe$_{0.72}$Pt$_{0.28}$ and Fe$_{0.68}$Pd$_{0.32}$ are shown in insets~(a) and~(b), respectively. This figure illustrates step 2 of the method described in Sec.~\ref{methods I}.}
\label{figure2}
\end{figure}

\begin{figure}
\includegraphics[width=8.5cm]{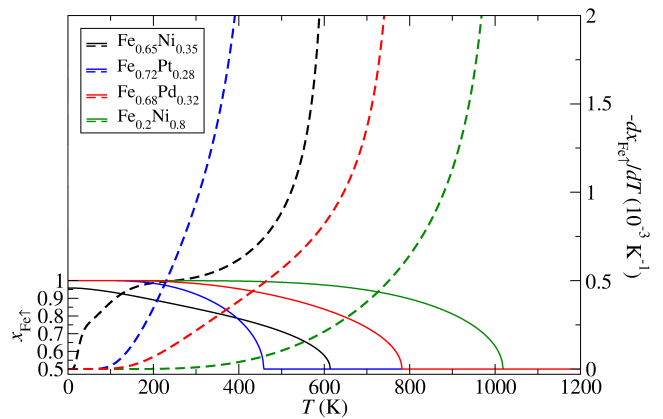}
\caption{Calculated magnetic properties associated with Fe sites as a function of temperature in fcc Fe$_{0.65}$Ni$_{0.35}$ (black lines), Fe$_{0.72}$Pt$_{0.28}$ (blue lines), Fe$_{0.68}$Pd$_{0.32}$ (red lines), and Fe$_{0.2}$Ni$_{0.8}$ (green lines). Fractions of Fe moments which point up correspond to solid lines, while demagnetization rates of Fe sites are indicated by dashed lines. This figure illustrates step 3 of the method.}
\label{figure3}
\end{figure}

\section{Computational methods}\label{methods I}

\begin{table*}[!t]
\caption{Theoretical and experimental results for fcc Fe-Ni, Fe-Pt, and Fe-Pd. Columns 2 and 3 display fractions of Fe moments which point up at several temperatures, according to the Ising model. Columns 4 and 5 show equilibrium lattice constants obtained by minimization of total energies (see steps 4 and 5 in Sec.~\ref{methods I}). Columns 7 and 8 show the result of minimizing free energies (see steps 6 and 7 in Sec.~\ref{methods I}). Note that, contrary to column 4,  the effect of zero-point lattice vibrations is included in the numbers in column 7. Columns 6 and 9 compare the effects of raising $x_{{\rm Fe}\uparrow}(T)$ from its value above the Curie temperature to its value at zero temperature on the calculated equilibrium lattice constants $a_{0}\big(x_{{\rm Fe}\uparrow}(T)\big)$ and $a\big(0, x_{{\rm Fe}\uparrow}(T)\big)$ for each alloy considered in this table. The rightmost column gives lattice constants measured at $4.2\,{\rm K}$ \protect\cite{acet94, matsui89}.}
\label{table1}
\begin{center}
\begin{tabular}{@{\hspace{0.1cm}} c c c c c c c c c c c c c c c c c c c c c c c c c}
\hline
\hline
\T \B \multirow{5}{*}{Alloy} & & & \multicolumn{19}{c}{Theory}  & & & Expt.\\
\cline{4-22} \cline{25-25}
\T \B & & & \multicolumn{3}{c}{$x_{{\rm Fe}\uparrow}(T)$}  & & & \multicolumn{3}{c}{$a_{0}\big(x_{{\rm Fe}\uparrow}(T)\big)$ (\AA)} & & & \multirow{3}{*}{$\Delta a_{0} / \Delta x_{{\rm Fe}\uparrow}$ (m\AA)} & & & \multicolumn{3}{c}{$a\big(0, x_{{\rm Fe}\uparrow}(T)\big)$ (\AA)} & & & \multirow{3}{*}{$\Delta a / \Delta x_{{\rm Fe}\uparrow}$ (m\AA)} & & & \multirow{3}{*}{$a$ (\AA)}\\
\cline{4-6} \cline{9-11} \cline{17-19}
\T \B & & & $T = 0$  & & $T > T_{\rm C}$ & & & $T = 0$ & & $T > T_{\rm C}$ & & & & & & $T = 0$ & & $T > T_{\rm C}$ & & & & & & \\
\hline 
Fe$_{0.65}$Ni$_{0.35}$ \T & & & 0.9576 & & 0.5 & & & 3.587 & & 3.553 & & & 74 & & & 3.595 & & 3.561 & & & 74 & & & 3.594 \\
Fe$_{0.6}$Ni$_{0.4}$ \T & & & 0.9909 & & 0.5 & & & 3.586 & & 3.555 & & & 63 & & & 3.594 & & 3.563 & & &  63 & & & 3.591 \\

Fe$_{0.55}$Ni$_{0.45}$ \T & & & 0.9992 & & 0.5 & & & 3.581 & & 3.555 & & & 52 & & & 3.59 & & 3.564 & & & 52 & & & 3.584 \\

Fe$_{0.5}$Ni$_{0.5}$  \T & & & 0.9999 & & 0.5 & & & 3.576 & & 3.555 & & & 42 & & & 3.585 & & 3.564 & & & 42 & & & 3.578 \\

Fe$_{0.45}$Ni$_{0.55}$ \T & & & 1 & & 0.5 & & & 3.571 & & 3.554 & & & 34 & & & 3.58 & & 3.564 & & & 32 & & & 3.57 \\
Fe$_{0.4}$Ni$_{0.6}$ \T & & & 1 & & 0.5 & & & 3.566 & & 3.554 & & & 24 & & & 3.574 & & 3.563 & & & 22 & & & 3.564 \\
Fe$_{0.35}$Ni$_{0.65}$ \T & & & 1 & & 0.5 & & & 3.56 & & 3.552 & & & 16 & & & 3.569 & & 3.561 & & & 16 & & & 3.558 \\
Fe$_{0.3}$Ni$_{0.7}$ \T & & & 1 & & 0.5 & & & 3.555 & & 3.548   & & & 14 & & & 3.564 & & 3.557 & & & 14 & & & 3.55 \\
Fe$_{0.25}$Ni$_{0.75}$ \T & & & 1 & & 0.5 & & & 3.55 & & 3.544  & & & 12 & & & 3.559 & & 3.554 & & & 10 & & & 3.545 \\
Fe$_{0.2}$Ni$_{0.8}$ \T & & & 1 & & 0.5 & & & 3.545 & & 3.54    & & & 10 & & & 3.554 & & 3.55 & & & 8 & & & 3.539 \\
Fe$_{0.72}$Pt$_{0.28}$ \T & & & 1 & & 0.5 & & & 3.775 & & 3.747 & & & 56 & & & 3.781 & & 3.753 & & & 56 & & & 3.752 \\
Fe$_{0.68}$Pd$_{0.32}$ \T \B & & & 1 & & 0.5 & & & 3.771 & & 3.753 & & & 36 & & & 3.779 & & 3.762 & & & 34 & & & 3.758 \\
\hline
\hline
\end{tabular}
\end{center}
\end{table*}

To study thermal expansion of Fe$_{1-x}${\it A}$_{x}$ over a broad temperature interval, we proceed as follows: 

1. We determine the five input parameters which are required for the Ising model of the M{\"u}ller-Hesse type \cite{muller83}:\ the magnitudes of local magnetic moments, $M_{\rm Fe}$ and $M_{A}$, and the three nearest-neighbor exchange constants, $J_{\rm FeFe}$, $J_{{\rm Fe}A}$, and $J_{AA}$. $M_{\rm Fe}$ and $M_{A}$ are calculated as the average magnetic moments on Fe sites and $A$ sites in an homogeneous ferromagnetic state by DFT. The method employed for $J_{AA}$ depends on the alloying element $A$. For {\it A}=Pt, Pd, $J_{AA}$ is taken to be zero:\ this rather crude assumption embodies the fact that bulk fcc metals Pt and Pd are both paramagnetic. On the other hand, for {\it A}=Ni, $J_{AA}$ is taken to be $J_{AA}$ of ferromagnetic fcc Ni \cite{footnote2, crangle62}. Furthermore, in all the cases, $J_{\rm FeFe}$ and $J_{{\rm Fe}A}$ are determined by fitting the calculated zero-temperature average magnetic moment per atom and the calculated Curie temperature to experimental data \cite{crangle62}. Before implementing this fitting procedure, we rescale the experimental Curie temperature by a factor of 1.23, to reflect the fact that our mean-field solution overestimates the exact Curie temperature by approximately 23\% \cite{skomski08}. 

2. Using the above-determined exchange parameters and magnetic moment magnitudes, we solve the disordered Ising model on the fcc lattice in the mean-field approximation. The disorder is included by using a separate mean field for sites with a different nearest-neighbor coordination. For example, we allow the mean field to be different on Fe sites with 9 Fe and 3 Ni nearest neighbors than on Fe sites with 10 Fe and 2 Ni nearest neighbors. Thus there are in total 26 mean fields, which are determined by a numerical solution of the appropriate self-consistency equations. From our solution, we calculate the average fraction of Fe moments with $n$ Fe nearest neighbors whose local moments are oriented in the `up' direction.

3. We estimate the fraction of Fe moments which point up, $x_{{\rm Fe} \uparrow}(T)$, using the results obtained from step 2. 

4. We perform DFT calculations of the total energy of the random alloy in a collinear magnetic state which reproduces the statistics of the local moments' orientations $x_{{\rm Fe}\uparrow}(T)$, $E\big(x_{{\rm Fe}\uparrow}(T), a\big)$, for various lattice constants $a$. Depending on the value of $x_{{\rm Fe}\uparrow}(T)$, the system is in homogeneous ferromagnetic states [case $x_{{\rm Fe}\uparrow}(T) = 1$], partially disordered local moment (PDLM) states [case $0.5 < x_{{\rm Fe}\uparrow}(T)<1$], or DLM states [case $x_{{\rm Fe}\uparrow}(T) = 0.5$]. In the two latter cases, up- and down-moments are randomly distributed on Fe sites. Total energies are calculated using the generalized gradient approximation (GGA) \cite{perdew96} and within the framework of the exact muffin-tin orbitals (EMTO) theory combined with the full charge density (FCD) technique \cite{vitos01}. As in recent theoretical studies on Fe-Ni \cite{ruban07, abrikosov07, ekholm10} and Fe-Pt \cite{khmelevskyi03, khmelevskyi03-PRB}, complete positional disorders of chemical species on fcc lattice sites and up- and down-moments on Fe sites are treated within the coherent potential approximation (CPA) \cite{vitos-abrikosov01}. Integration in the irreducible wedge of the Brillouin zone is carried out over several thousands of $\mathbf{k}$-points generated according to the Monkhorst-Pack scheme \cite{monkhorst76}. 

5. We fit the results of step 4 with a Morse function. The parameters of the fit give the equilibrium lattice constant, $a_{0}\big(x_{{\rm Fe}\uparrow}(T)\big)$, the bulk modulus, $B_{0}\big(x_{{\rm Fe}\uparrow}(T)\big)$, and the Gr{\"u}neisen constant, $\gamma_{0}\big(x_{{\rm Fe}\uparrow}(T)\big)$. 

6. For each lattice constant chosen in step 4, we add to the total energy $E\big(x_{{\rm Fe}\uparrow}(T), a\big)$ a vibrational free energy contribution to the Helmholtz free energy, $F_{\rm vib}(T, x_{{\rm Fe}\uparrow}(T), a)$. The latter is estimated within the Debye-Gr{\"u}neisen model from the outputs of step 5, $a_{0}\big(x_{{\rm Fe}\uparrow}(T)\big)$, $B_{0}\big(x_{{\rm Fe}\uparrow}(T)\big)$, and $\gamma_{0}\big(x_{{\rm Fe}\uparrow}(T)\big)$. The sum of the two terms mentioned above can be written as
\begin{eqnarray}
& F\big(T, x_{{\rm Fe}\uparrow}(T), a\big) = E\big(x_{{\rm Fe}\uparrow}(0), a\big) & \nonumber \\
& + \,F_{\rm mag}\big(x_{{\rm Fe}\uparrow}(T), a\big) + \,F_{\rm vib}(T, x_{{\rm Fe}\uparrow}(T), a), & \label{eq2}
\end{eqnarray}
where
\begin{eqnarray}
& F_{\rm mag}\big(x_{{\rm Fe}\uparrow}(T), a\big) = E\big(x_{{\rm Fe}\uparrow}(T), a\big) - \,E\big(x_{{\rm Fe}\uparrow}(0), a\big). & \label{eq3}
\end{eqnarray}

7. We minimize the contribution to the Helmholtz free energy~(\ref{eq2}) with respect to $a$ to obtain the equilibrium lattice spacing $a\big(T, x_{{\rm Fe}\uparrow}(T)\big)$. 

8. We repeat steps 2 to 7 with different temperatures. Subsequently, we apply a cubic-spline interpolation procedure. 

9. We evaluate the thermal expansion coefficient
\be
\alpha(T) = \lim_{\delta T \rightarrow 0} \frac{a\big(T + \,\delta T, x_{{\rm Fe}\uparrow}(T + \,\delta T)\big) - \,a\big(T, x_{{\rm Fe}\uparrow}(T)\big)}{a\big(T, x_{{\rm Fe}\uparrow}(T)\big)\,\delta T}, \label{alpha}
\ee
for the dense set of data from step 8.

\section{Results and analysis}\label{results I}

According to experiments, ferromagnetic Fe$_{1-x}$Ni$_{x}$ with $x = 0.35, 0.4,\ldots, 0.8$, Fe$_{0.72}$Pt$_{0.28}$, and Fe$_{0.68}$Pd$_{0.32}$ exhibit a wide variety of thermal behavior, some showing the Invar effect \cite{matsui80, tanji71, sumiyama79} and others presenting thermal expansion similar to that of paramagnetic alloys \cite{tanji71}. For this reason, they represent an attractive choice for testing the general approach presented in Sec.~\ref{methods I}. 

We begin the test by considering the input parameters of the Ising model. The average magnetic moments on each type of site in the homogeneous ferromagnetic binary alloys are calculated at zero temperature by means of the EMTO method. Our calculated moments on Fe sites cover the range from $2.63\,\mu_{\rm B}$ for Fe$_{0.65}$Ni$_{0.35}$ to $2.89\,\mu_{\rm B}$ for Fe$_{0.68}$Pd$_{0.32}$; the moments on Ni, Pt, and Pd sites are found to span the interval from $0.3\,\mu_{\rm B}$ for Fe$_{0.68}$Pd$_{0.32}$ to $0.64\,\mu_{\rm B}$ for Fe$_{0.2}$Ni$_{0.8}$. All these results yield fair agreement with available DFT data \cite{khmelevskyi03, khmelevskyi03-PRB, ruban07, ekholm10, liot09} and experimental measurements \cite{nishi74, capolaretti80}. 

\begin{figure}
\includegraphics[width=8cm]{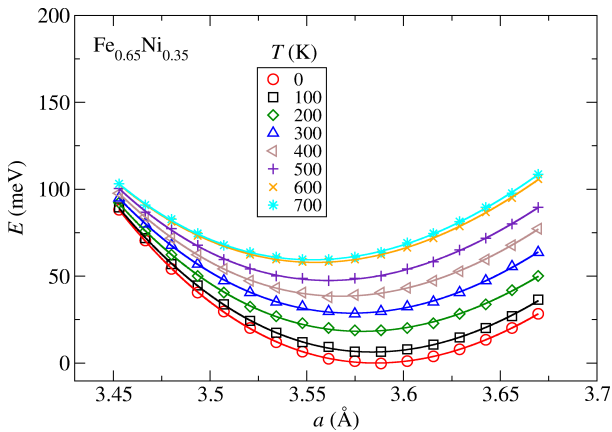}
\caption{Calculated total energy per atom of fcc Fe$_{0.65}$Ni$_{0.35}$ in a collinear magnetic state $E\big(x_{{\rm Fe}\uparrow}(T), a\big)$ plotted against lattice constant for various temperatures. This figure illustrates step 4 of the method.}
\label{figure4}
\end{figure}

\begin{figure}
\includegraphics[width=8cm]{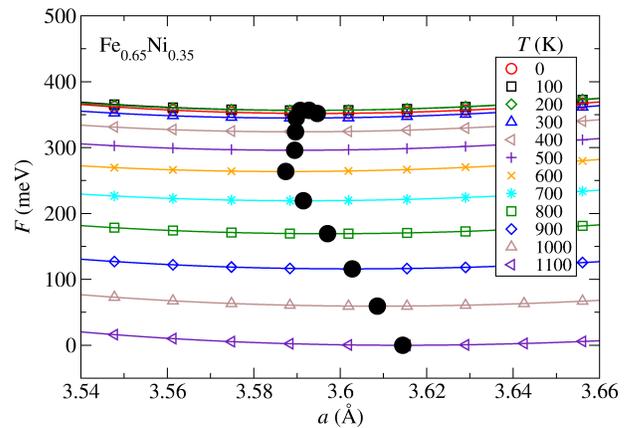}
\caption{Free energy of fcc Fe$_{0.65}$Ni$_{0.35}$ in a collinear magnetic state $F\big(T, x_{{\rm Fe}\uparrow}(T), a\big)$ as determined from Eq.~(\ref{eq2}) versus lattice constant for several temperatures. Black filled circles depict equilibrium lattice parameters.}
\label{figure5}
\end{figure}

\begin{figure}
\includegraphics[width=8cm]{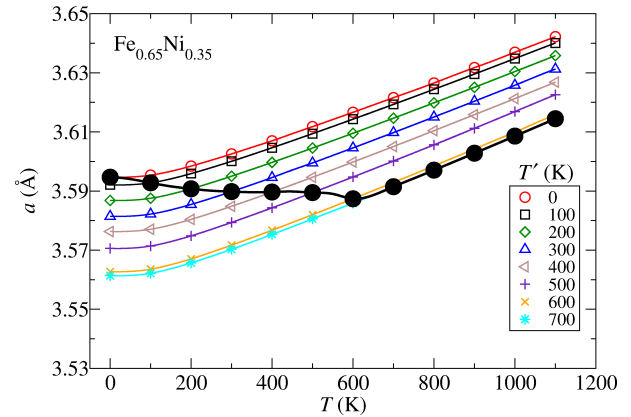}
\caption{Estimated equilibrium lattice parameter of fcc Fe$_{0.65}$Ni$_{0.35}$ in a collinear magnetic state $a\big(T, x_{{\rm Fe}\uparrow}(T')\big)$ plotted against temperature $T$ for various values of temperature $T'$. The same symbol as in Fig.~\ref{figure5} marks the calculated equilibrium lattice parameter $a\big(T, x_{{\rm Fe}\uparrow}(T)\big)$ for $T = 0, 100,\ldots, 1100\,{\rm K}$. The thick black solid line results from applying a cubic-spline interpolation scheme to a data set which is almost twice as large as the number of black filled circles. From this curve, we obtain thermal expansion coefficients [see Fig.~\ref{figure7}(a)].}
\label{figure6}
\end{figure}

Fig.~\ref{figure1} displays data for Fe-Ni. While the inset compares calculated Curie temperatures to rescaled experimental findings, the main panel compares calculated average magnetic moments per atom at zero temperature to measurements. The quantitative agreement of the numerical results with the corresponding experimental observations is achieved by varying $J_{\rm FeFe}$ and $J_{{\rm FeNi}}$ for each considered Ni concentration. The fitting procedure leads to an enhancement of the Fe-Fe exchange parameter as the Ni content increases, from a negative value for $x=0.35$ to a positive value for $x=0.8$. This behavior is consistent with Monte Carlo simulations \cite{rancourt96}. Interestingly, when applied to Fe$_{0.72}$Pt$_{0.28}$ and Fe$_{0.68}$Pd$_{0.32}$, the fitting procedure described in step 1 in Sec.~\ref{methods I} gives ferromagnetic coupling between moments on neighboring Fe sites ($J_{\rm FeFe} > 0$).

Estimated fractions of all Fe moments with $n$ Fe first neighbors which point up are presented in Fig.~\ref{figure2} for $n = 0, 1, \ldots, 12$; calculated fractions of Fe moments which point up are plotted against temperature in Fig.~\ref{figure3}. Fe$_{0.72}$Pt$_{0.28}$ and Fe$_{0.68}$Pd$_{0.32}$ are found to exhibit homogeneous ferromagnetism at zero temperature, while the magnetic structure of Fe$_{0.65}$Ni$_{0.35}$ appears to consist of 97\% of up-moments and 3\% of down-moments. These results reproduce available experimental observations \cite{abd87, nakamura79}. For Fe$_{0.72}$Pt$_{0.28}$, Fe$_{0.68}$Pd$_{0.32}$, and Fe$_{0.65}$Ni$_{0.35}$, the calculations give $x_{{\rm Fe}\uparrow}(T) = 1, 1, 0.9576$, respectively, at zero temperature. The corresponding values drop by 13, 15, and 22\% at the reduced temperature $T/T_{\rm C} = 0.75$. The fraction of Fe moments which point up in the Fe-Ni alloy is therefore predicted to significantly underestimate that of the Fe-Pt and Fe-Pd alloys not only at zero temperature but also near the Curie temperature. Analysis of Fig.~\ref{figure2} provides insight into how up- and down-moments are distributed among Fe sites for various temperatures. In the Fe-Ni alloy at zero temperature, down-moments are found to reside exclusively on Fe sites with 11 and 12 Fe nearest neighbors. This picture is consistent with recent density-functional total-energy calculations performed within the local spin-density approximation (LSDA) at the experimental lattice spacing of $3.59\,{\rm \AA}$ \cite{ruban07}. Perhaps more surprisingly, Fig.~\ref{figure2} reveals that the distribution of up-moments on Fe sites for any reduced temperature in the range 0-0.75, more closely resembles a random distribution in the Fe-Pt alloy. Accordingly, we expect the methodology introduced in Sec.~\ref{methods I} to produce more accurate thermal expansion coefficients for Fe-Pt than for Fe-Ni.

\begin{figure}[t]
\includegraphics[width=8.1cm]{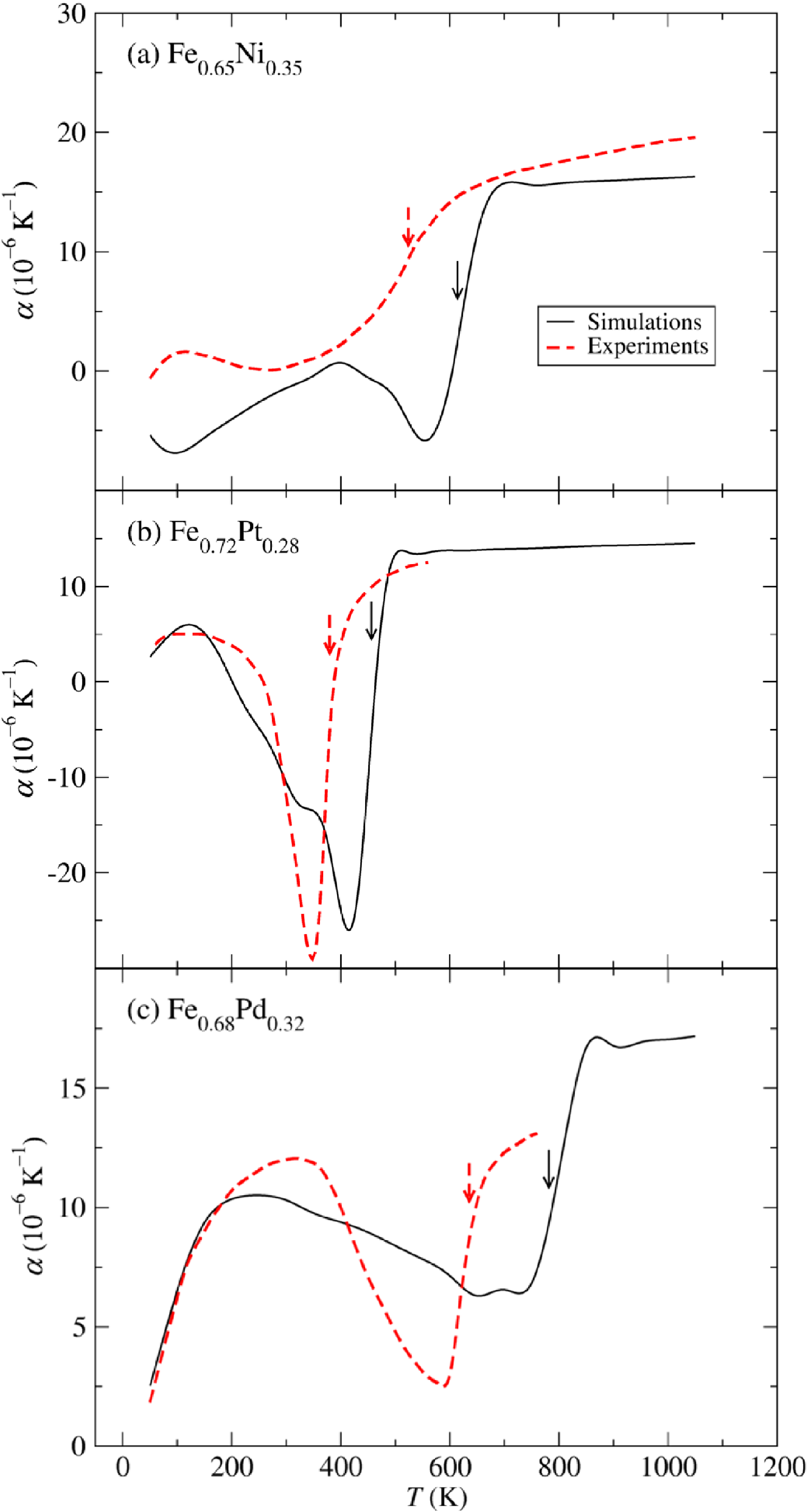}
\caption{Linear thermal expansion coefficients of fcc ferromagnets plotted as a function of temperature. Panel~(a): Fe$_{0.65}$Ni$_{0.35}$. Panel~(b): Fe$_{0.72}$Pt$_{0.28}$. Panel~(c): Fe$_{0.68}$Pd$_{0.32}$. Solid lines show results obtained following the procedure described in Sec.~\ref{methods I}. Dashed lines correspond to experimental data \cite{matsui78, sumiyama79, matsui80}. Vertical arrows indicate Curie temperatures. Our observation that these materials all display the Invar effect perfectly matches experimental findings. In addition, the theory correctly predicts the overall trends in $\alpha$ versus $T$ for Fe$_{0.72}$Pt$_{0.28}$ and Fe$_{0.68}$Pd$_{0.32}$.}
\label{figure7}
\end{figure}

In Fig.~\ref{figure4}, calculated total energy of Fe$_{0.65}$Ni$_{0.35}$ in a collinear magnetic state $E\big(x_{{\rm Fe}\uparrow}(T), a\big)$ is plotted as a function of lattice constant for temperature intervals of $100\,{\rm K}$. The estimated values for the equilibrium lattice constant $a_{0}\big(x_{{\rm Fe}\uparrow}(T)\big)$ at zero temperature and above the Curie temperature are reported in Table~\ref{table1} along with those of other compounds. The curves in Fig.~\ref{figure4} are analyzed in light of Fig.~\ref{figure3}: The equilibrium lattice constant $a_{0}\big(x_{{\rm Fe}\uparrow}(T)\big)$ shifts continuously towards larger values with increasing the fraction of Fe moments which point up in the system. This confirms expectations based on Refs.~\onlinecite{abrikosov07} and~\onlinecite{khmelevskyi03}. Actually, a behavior similar to that observed in Fe$_{0.65}$Ni$_{0.35}$ is seen in each of the other systems investigated. To get a rough estimate of the effect in each alloy, we evaluate the ratio $\Delta a_{0} / \Delta x_{{\rm Fe}\uparrow} = \big[ a_{0}\big(x_{{\rm Fe}\uparrow}(0)\big) - a_{0}(0.5) \big] / [x_{{\rm Fe}\uparrow}(0) - 0.5]$. The estimated values are displayed in Table~\ref{table1}, indicating that the dependence on the fraction of Fe moments which point up is more pronounced in the Fe-rich alloys Fe$_{0.65}$Ni$_{0.35}$, Fe$_{0.72}$Pt$_{0.28}$, and Fe$_{0.68}$Pd$_{0.32}$ than in the Ni-rich alloy Fe$_{0.2}$Ni$_{0.8}$. 
 
The result of applying the sixth, seventh, and eighth steps of the procedure is shown in Figs.~\ref{figure5} and~\ref{figure6} for Fe$_{0.65}$Ni$_{0.35}$. In Fig.~\ref{figure5}, the total energy $E\big(x_{{\rm Fe}\uparrow}(T), a\big)$ is added to the vibrational free energy $F_{\rm vib}(T, x_{{\rm Fe}\uparrow}(T), a)$ and their sum $F(T, x_{{\rm Fe}\uparrow}(T), a)$ is plotted over a narrow range of lattice constants for several temperatures below and above the Curie temperature of $614\,{\rm K}$. The positions of the minima in free-energy curves are marked by black filled circles. Results are reported in Fig.~\ref{figure6}. In accordance with experimental observations \cite{oomi81}, it is found that the equilibrium lattice constant in the range 0-$T_{\rm C}$ clearly displays a deviation from the monotonically increasing behavior seen above the critical temperature.

While this paper focuses on thermal expansion, it is interesting to test whether our simulation properly models departures from Vegard's law for Fe-Ni alloys (see, e.g., Ref.~\onlinecite{acet94}). Similarly to the Invar effect, characteristic negative deviations from linear behavior in the lattice constant at very low temperature still await for a complete understanding. As can be seen in Table~\ref{table1}, the calculated $a\big(0, x_{{\rm Fe}\uparrow}(0)\big)$ agrees closely with a measured lattice constant \cite{acet94, matsui89} irrespective of the chemical composition of the considered material. With the exception of Fe$_{0.65}$Ni$_{0.35}$, our data for the Fe-Ni series are fitted linearly as $a\big(0, x_{{\rm Fe}\uparrow}(0)\big)$ versus $x$. The resulting relative deviation from Vegard's law at $x = 0.35$ amounts to -0.14\%, which is in quantitative agreement with the tiny experimental value of -0.08\%. 

We now turn to the central question of how well the model captures the rich variety of thermal expansion phenomena observed in ferromagnetic Fe$_{1-x}$Ni$_{x}$ with $x = 0.35, 0.4,\ldots, 0.8$, Fe$_{0.72}$Pt$_{0.28}$, and Fe$_{0.68}$Pd$_{0.32}$. 

\begin{figure}
\includegraphics[width=8cm]{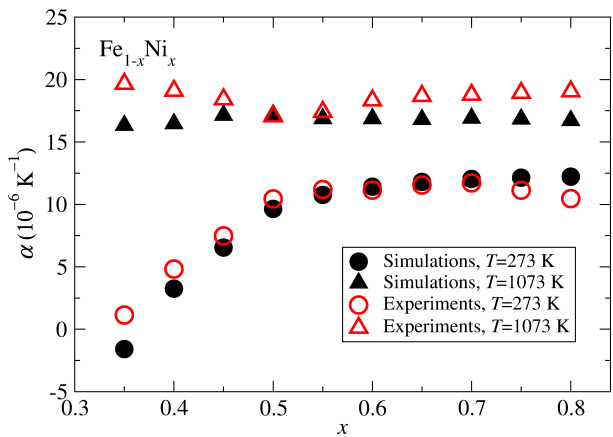}
\caption{Linear thermal expansion coefficient of fcc Fe$_{1-x}$Ni$_{x}$ versus nickel atomic concentration, according to the method presented in Sec.~\ref{methods I} (filled symbols) and experiments \cite{tanji71} (open symbols). Circles and triangles show results for $T = 273\,{\rm K}$ and $T = 1073\,{\rm K}$, respectively. Our approach reliably reproduces the significant reduction of the thermal expansion coefficient of Fe$_{1-x}$Ni$_{x}$ as $x$ decreases from 0.55 to 0.35 near room temperature, which was discovered by Guillaume \cite{guillaume67}.}
\label{figure8}
\end{figure}

Figs.~\ref{figure7} and~\ref{figure8} provide a comparison of computed and experimentally-determined \cite{matsui78, sumiyama79, matsui80} linear thermal expansion coefficients. While Fig.~\ref{figure7} illustrates the temperature dependence of calculated and measured structural properties of Fe$_{0.65}$Ni$_{0.35}$ [panel~(a)], Fe$_{0.72}$Pt$_{0.28}$ [panel~(b)], and Fe$_{0.68}$Pd$_{0.32}$ [panel~(c)], Fig.~\ref{figure8} shows how results obtained for Fe$_{1-x}$Ni$_{x}$ vary with nickel concentration at fixed temperature.

The model performs well, at least at a qualitative level, throughout the whole spectrum of test compounds. Indeed, our observation that Fe$_{0.65}$Ni$_{0.35}$, Fe$_{0.72}$Pt$_{0.28}$, and Fe$_{0.68}$Pd$_{0.32}$ all display anomalously small thermal expansion coefficients over broad temperature ranges (i.e., the Invar effect) perfectly matches experimental findings. In addition, the theory correctly predicts the overall trends in $\alpha$ versus $T$ for Fe$_{0.72}$Pt$_{0.28}$ and Fe$_{0.68}$Pd$_{0.32}$. Even the significant reduction of the thermal expansion coefficient of Fe$_{1-x}$Ni$_{x}$ as $x$ decreases from 0.55 to 0.35 near room temperature, which was discovered by the Nobel prize winner Guillaume \cite{guillaume67}, is reliably reproduced.


\section{Conclusion} \label{conclusion}

To investigate theoretically the Invar effect in ferromagnetic disordered fcc Fe-{\it A} with {\it A}=Ni, Pt, Pd, a computationally efficient scheme inspired by previous work \cite{khmelevskyi03, moruzzi88} has been designed. The procedure can be divided into two stages: study of magnetism and calculations of structural properties. In the first stage, an Ising model is considered and fractions of Fe moments which point up as a function of temperature are determined. In the second stage, DFT calculations are performed to evaluate free energies of alloys in PDLM and DLM states as a function of lattice constant for various temperatures. It is worth emphasizing that neither noncollinear magnetism \cite{vanschilfgaarde99} nor partial chemical ordering \cite{crisan02} are explicitly taken into account at any stage.

Extensive tests of the approach have been carried out by comparing simulation results for thermal expansion coefficients of Fe$_{1-x}$Ni$_{x}$ with $x = 0.35, 0.4,\ldots, 0.8$, Fe$_{0.72}$Pt$_{0.28}$, and Fe$_{0.68}$Pd$_{0.32}$ with measurements. Despite a number of approximations (e.g., neglect of static ionic displacements \cite{liot06, liot09}), the scheme has been found to perform well, at least qualitatively, throughout the whole spectrum of test compounds.

As a result of the overall qualitative agreement between theory and experiment, it appears that the Invar effect in Fe-Ni can be investigated within the same computational framework as Fe-Pt and Fe-Pd. This represents significant progress compared to previous schemes that incorporate DFT calculations.  

In addition, tests results provide evidence that the methodology captures the essential physics of the Invar effect. For this reason, the present work is currently being extended to achieve a better understanding of the physical mechanism behind the remarkable phenomenon \cite{liot11}. 

\begin{acknowledgments}
The interest and support of I. A. Abrikosov are gratefully acknowledged. F. L. thanks B. Alling, M. Ekholm and P. Steneteg for helping him with calculations. This work was supported by grants from the Swedish Research Council (VR), the Swedish Foundation for Strategic Research (SSF), the G{\"o}ran Gustafsson Foundation for Research in Natural Sciences and Medicine, the EPSRC (UK), the Scottish Universities Physics Alliance, and the HPC-Europa project.
\end{acknowledgments}


\begin{thebibliography}{31}

\bibitem{kamihara08}
Y. Kamihara, T. Watanabe, M. Hirano, and H. Hosono, J. Am. Chem. Soc. {\bf130}, 3296 (2008).

\bibitem{brando08}
M. Brando, W. J. Duncan, D. Moroni-Klementowicz, C. Albrecht, D. Gr\"{u}ner, R. Ballou, and F. M. Grosche, Phys. Rev. Lett. \textbf{101}, 026401 (2008).

\bibitem{wassermann90} 
E. F. Wassermann, in {\it Ferromagnetic Materials}, edited by K. H. J. Buschow and E. P. Wohlfarth (North-Holland, Amsterdam, 1990).

\bibitem{guillaume97} 
C. E. Guillaume, C.R. Acad. Sci. {\bf125}, 235 (1897).

\bibitem{matsui80}
M. Matsui, T. Shimizu, H. Yamada, and K. Adachi, J. Magn. Magn. Mater. \textbf{15}, 1201 (1980).

\bibitem{wassermann98}
E. F. Wassermann, B. Rellinghaus, T. Roessel, J. K\"{a}stner, and W. Pepperhoff, Eur. Phys. J. B \textbf{5}, 361 (1998).

\bibitem{khmelevskyi05}
S. Khmelevskyi, A. V. Ruban, Y. Kakehashi, P. Mohn, and B. Johansson, Phys. Rev. B \textbf{72}, 064510 (2005).

\bibitem{ruban07}
A. V. Ruban, S. Khmelevskyi, P. Mohn, and B. Johansson, Phys. Rev. B \textbf{76}, 014420 (2007).

\bibitem{abrikosov07}
I. A. Abrikosov, A. E. Kissavos, F. Liot, B. Alling, S. I. Simak, O. Peil, and A. V. Ruban, Phys. Rev. B \textbf{76}, 014434 (2007).

\bibitem{khmelevskyi03}
S. Khmelevskyi, I. Turek, and P. Mohn, Phys. Rev. Lett. \textbf{91}, 037201 (2003).

\bibitem{footnote1}
A first-principles method using the DLM approach was first introduced in the 1980s \cite{gyorffy85}, and applied to the Invar problem in the 1990s \cite{johnson90}. Such a computational technique has since been utilized extensively.

\bibitem{gyorffy85}
B. L. Gyorffy, A. J. Pindor, J. Staunton, G. M. Stocks, and H. Winter, J. Phys. F: Met. Phys. {\bf15}, 1337 (1985).

\bibitem{johnson90}
D. D. Johnson, F. J. Pinski, J. B. Staunton, B. L. Gyorffy, and G. M. Stocks, in {\it Physical Metallurgy of Controlled Expansion Invar-Type Alloys}, edited by K. C. Russel and D. F. Smith (TMS, Warrendale, PA, 1990).

\bibitem{moruzzi88}
V. L. Moruzzi, J. F. Janak, and K. Schwarz, Phys. Rev. B \textbf{37}, 790 (1988).

\bibitem{herper99}
H. C. Herper, E. Hoffmann, and P. Entel, Phys. Rev. B \textbf{60}, 3839 (1999).

\bibitem{muller83}
J. B. M\"{u}ller and J. Hesse, Z. Phys. B \textbf{54}, 35 (1983).

\bibitem{footnote2}
For pure fcc Ni, $M_{\rm Ni}$ is determined by DFT calculations in a ferromagnetic state and $J_{\rm NiNi}$ by fitting Ising model results to experimental data \cite{crangle62}.

\bibitem{crangle62}
J. Crangle and G. C. Hallam, Proc. R. Soc. Lond. A {\bf272}, 119 (1963).

\bibitem{skomski08}
R. Skomski, in {\it Simple Models of Magnetism} (Oxford University Press, 2008).

\bibitem{perdew96} 
J. P. Perdew, K. Burke, and M. Ernzerhof, Phys. Rev. Lett. {\bf 77}, 3865 (1996).

\bibitem{vitos01}

L. Vitos, Phys. Rev. B \textbf{64}, 014107 (2001).

\bibitem{ekholm10}
M. Ekholm, H. Zapolsky, A. V. Ruban, I. Vernyhora, D. Ledue, and I. A. Abrikosov, Phys. Rev. Lett. {\bf 105}, 167208 (2010).

\bibitem{khmelevskyi03-PRB}
S. Khmelevskyi and P. Mohn, Phys. Rev. B \textbf{68}, 214412 (2003).


\bibitem{vitos-abrikosov01}
L. Vitos, I. A. Abrikosov, and B. Johansson, Phys. Rev. Lett. \textbf{87}, 156401 (2001).

\bibitem{monkhorst76}
H. J. Monkhorst and J. D. Pack, Phys. Rev. B \textbf{13}, 5188 (1976).


\bibitem{tanji71}

Y. Tanji, J. Phys. Soc. Jpn. \textbf{31}, 1366 (1971).

\bibitem{sumiyama79}
K. Sumiyama, M. Shiga, M. Morioka, and Y. Nakamura, J. Phys. F: Met. Phys. \textbf{9}, 1665 (1979).

\bibitem{liot09}
F. Liot and I. A. Abrikosov, Phys. Rev. B \textbf{79}, 014202 (2009).

\bibitem{nishi74}
M. Nishi, Y. Nakai, and N. Kunitomi, J. Phys. Soc. Jpn. \textbf{37}, 570 (1974).

\bibitem{capolaretti80}
O. Caporaletti and G. M. Graham, J. Magn. Magn. Mater. \textbf{22}, 25 (1980).

\bibitem{rancourt96}
D. G. Rancourt and M.-Z. Dang, Phys. Rev. B \textbf{54}, 12225 (1996).

\bibitem{abd87}
M. M. Abd-Elmeguid, U. Hobuss, H. Micklitz, B. Huck, and J. Hesse, Phys. Rev. B \textbf{35}, 4796 (1987).

\bibitem{nakamura79}
Y. Nakamura, K. Sumiyama, and M. Shiga, J. Magn. Magn. Mater. \textbf{12}, 127 (1979).

\bibitem{oomi81}
G. Oomi and N. M\={o}ri, J. Phys. Soc. Jpn. \textbf{50}, 2924 (1981).

\bibitem{acet94}
M. Acet, H. Z\"{a}hres, E. F. Wassermann, and W. Pepperhoff, Phys. Rev. B \textbf{49}, 6012 (1994).

\bibitem{matsui89}
M. Matsui and K. Adachi, Physica B \textbf{161}, 53 (1989).

\bibitem{matsui78}
M. Matsui and S. Chikazumi, J. Phys. Soc. Jpn. \textbf{45}, 458 (1978).

\bibitem{guillaume67}
C. E. Guillaume, in {\it Nobel Lectures in Physics 1901-1921} (Elsevier, Amsterdam, 1967).

\bibitem{vanschilfgaarde99}
M. van Schilfgaarde, I. A. Abrikosov, and B. Johansson, Nature \textbf{400}, 46 (1999).

\bibitem{crisan02}
V. Crisan, P. Entel, H. Ebert, H. Akai, D. D. Johnson, and J. B. Staunton, Phys. Rev. B \textbf{66}, 014416 (2002).

\bibitem{liot06}
F. Liot, S. I. Simak, and I. A. Abrikosov, J. Appl. Phys. \textbf{99}, 08P906 (2006).

\bibitem{liot11}
F. Liot and C. A. Hooley, e-print arXiv:0912.0215v4 (unpublished).

\end{thebibliography}

\end{document}